\documentclass[11pt,letterpaper,english,aps]{revtex4}
\usepackage{times}
\usepackage[T1]{fontenc}
\usepackage[latin1]{inputenc}
\usepackage{subfigure}
\usepackage{float}
\usepackage{amsmath}
\usepackage{graphicx}

\makeatletter
\usepackage{babel}
\makeatother
\begin{document}

\title{Self-organized metal nanostructures through laser driven thermocapillary convection}

\author{C. Favazza, J. Trice, %
\thanks{
\email{ramkik@wuphys.wustl.edu}\maketitle
}R. Kalyanaraman}

\affiliation{Department of Physics, Washington University in St. Louis, MO 63130}

\affiliation{Center for Materials Innovation, Washington University in St. Louis, MO 63130}

\author{%
\thanks{
\email{suresh@che.wustl.edu}\maketitle
}R. Sureshkumar}

\affiliation{Department of Energy, Environmental, and Chemical Engineering, Washington University in St.
Louis, MO 63130}

\affiliation{Center for Materials Innovation, Washington University in St. Louis, MO 63130}

\begin{abstract}
When ultrathin metal films are subjected to multiple cycles of rapid melting and resolidification
by a ns pulsed laser, spatially correlated interfacial nanostructures can result from a competition
among several possible thin film self-organizing processes. Here we investigate self-organization
and the ensuing length scales when Co films ($1-8\, nm$ thick) on SiO$_{\text{2}}$ surfaces
are repeatedly and rapidly melted by non-uniform (interference) laser irradiation. Pattern evolution
produces nanowires, which eventually break-up into nanoparticles exhibiting spatial order in
the nearest neighbor spacing, $\lambda_{NN2}$. For films of thickness $h_{0}>2\, nm$, $\lambda_{NN2}\propto h_{0}^{1/2}$
while the particle radius varies as $r_{p2}\propto h_{0}^{1/2}$. This scaling behavior is consistent
with pattern formation by thermocapillary flow and a Rayleigh-like instability. For $h_{0}\leq2\, nm$,
a hydrodynamic instability of a spinodally unstable film leads to the formation of nanoparticles. 
\end{abstract}
\maketitle
Metal nanostructures, possessing spatial order and size uniformity are attractive for a variety
of applications including: optoelectronics, plasmonics, chemical and biomedical sensing, and
catalytic devices \cite{quinten98-a,garcia06c,Shipway,Xia,Raty}. In recent years there has been
tremendous research emphasis on finding self-organization strategies to make strongly correlated
metal and semiconductor nanostructures. In the area of thin film pattern formation, epitaxial
strain driven self-organization in crystallographic systems has been somewhat successful \cite{Brune,Kwak,Teichert,Goswami}.
However, this strategy of exploiting epitaxial strain cannot be applied to materials on  important
amorphous surfaces, such as SiO$_{\text{2}}$. One possibility for self-organization on amorphous
surfaces is via dewetting instabilities, such as those observed in liquid polymer films on various
inert surfaces, in which  preferred length scales appear when destabilizing long range attractive
intermolecular forces overcome stabilizing surface tension forces \cite{vrij,vrij2,deGennes}.
 We have previously shown that multiple pulses of single beam ns laser irradiation of Co films
with thickness $1\le h_{0}\le8\, nm$ on SiO$_{\text{2}}$ surfaces demonstrate self-organization
by a thin film hydrodynamic (TFH) instability with length scales characterized by the metal surface
tension ($\gamma$),  the dispersion force characterized by the Hamaker constant $A$ and film
thickness $h_{0}$ \cite{favazza06d}. The ensuing pattern had short range spatial order (SRO)
in the nearest-neighbor ($NN$) particle spacing $\lambda_{NN1}$, which scaled with thickness
as $h_{0}^{2}$, while the particle radius varied as $r_{p1}\sim h_{0}^{5/3}$.  Recently we
have also provided preliminary evidence that 2-beam laser interference irradiation can give rise
to \emph{quasi-2D} ordering of the nanoparticles with both long range order (LRO) and SRO \cite{favazza06c}.
However the evolution of the 2-beam pattern from the flat film to nanoparticle state was not
investigated, and the mechanisms for spatial ordering were not identified. Here we report detailed
experiments of 2-beam interference irradiation of Co on SiO$_{\text{2}}$ as a function of film
thickness for $1\leq h_{0}\leq8\, nm$ that demonstrate that the pattern evolution and resulting
length scales are consequences of different mechanisms, including the TFH instability, thermocapillary
($TC$) flow and a Rayleigh-like instability, based on the timescales of these processes.  One
can access a particular mechanism via the choice of film thickness and/or thermophysical parameters
such as $\gamma$, $A$ and the temperature gradient.

Cobalt films ranging in thickness from $\sim1-8\, nm$ were deposited by e-beam evaporation onto
optically smooth SiO$_{\text{2}}$ surfaces in vacuum. Specific details of the experiments are
published elsewhere \cite{favazza06c,Favazza06b}. In brief, following deposition, the Co films
were irradiated in vacuum with a $266\, nm$ pulsed laser operating at a repetition rate of $50\, Hz$
with a pulse time $\tau\sim9\, ns$ by single beam \cite{favazza06d} or 2-beam interference
irradiation. In the single beam case, films were exposed to an unfocused beam at normal incidence.
For the 2-beam condition, an interference angle of $\sim45^{\circ}$ generated a periodic laser
intensity profile whose contrast was maximized by adjusting the energy density of the off-normal
beam with a $750\, mm$ lens positioned $300\, mm$ away from the sample at angle of $\sim45^{\circ}$.
In both instances, Co films were irradiated for similar times as measured by the number of laser
pulses, $n$, for $5\le n\le10,500$. Also, comparable laser energy densities  were used for
both types of irradiation, with the energy density, $E_{laser}$ between $E_{Threshold}<60\le E_{laser}\le200\, mJ$
\cite{Trice06a}. The resultant morphological evolution was investigated as a function of $n$
and was examined and characterized using a Hitachi S-4500 scanning electron microscope (SEM).
It was also verified that all the films studied had minimal evaporation, as confirmed by performing
energy dispersive X-ray spectrometry measurements of the Co concentration after irradiation.

As detailed in prior studies for single beam irradiation \cite{favazza06d,Trice06a}, the film
shows several distinct patterns enroute to a final robust state of nanoparticles characterized
by a spatially correlated nearest neighbor ($NN$) spacing. Here we briefly summarize the trends
for the purpose of comparison with the pattern evolution under 2-beam interference irradiation.
More detailed analysis the single beam data can be found elsewhere \cite{favazza06d,Trice06a}.
At the early stages ($n=10$)  regular holes with a characteristic diameter form, followed by
a cellular network of polygonal structures. Eventually, the polygonal networks evolve into nanoparticles,
which form predominantly at the vertices of the polygons. An important property of the evolution
is that the patterns possessed a characteristic length scale at every stage. In Fig. \ref{fig:LengthscalePlots}(a),
the scaling behavior  of the spatial correlation in the final stable nanoparticle state is shown
for varying initial film thicknesses. The observed trend with $h_{0}$ was in agreement with
classical linear TFH dewetting theory \cite{vrij,vrij2} in which: \begin{equation}
\Lambda_{TFH}=\sqrt{\frac{16\pi^{3}\gamma}{A}}h_{o}^{2}\label{eq:SpinLengthscale}\end{equation}
 where $\Lambda_{TFH}$ represents the average $NN$ nanoparticle spacing, $\lambda_{NN1}$,
$A=1.4\times10^{-18}J$ is the experimentally estimated Hamaker constant for the ${SiO}_{\text{2}}/Co/Vacuum$
system , $\gamma=1.88\,\frac{mJ}{m^{2}}$  is the Co liquid surface tension and $h_{0}$ is the
film thickness \cite{vrij,vrij2}. Furthermore, volume conservation implied that the nanoparticles
will have a radius that will vary with the initial film thickness $h_{0}$ as:\begin{equation}
r_{p1}=\sqrt[3]{\frac{24\pi^{3}\gamma}{f(\theta)A}}h_{o}^{\frac{5}{3}}\label{eq:PartRadius}\end{equation}
 where $f(\theta)$ accounts for the contact angle of the nanoparticles \cite{favazza06d}. We
experimentally observed a monomodal particle size distribution with the average radius shown
in Fig. \ref{fig:LengthscalePlots}(b) and a trend consistent with the above theoretical prediction
from linear TFH theory. 

The results of 2-beam irradiation were qualitatively different from single beam irradiation.
In Fig. \ref{fig:PatternProgression}(a-c), the typical pattern evolution for films with $h_{0}>2\, nm$
is shown as a function of $n$. In this case, the early stages are comprised of spatially periodic
film rupture at length scales comparable to the interference spacing. Longer irradiation yielded
the formation of long, cylindrical-like {}``nanowires'' and continued irradiation resulted
in the break-up of these nanowires into particles. The final particle state is characterized
by a quasi-2D, comprised of the LRO due to periodic laser intensity and SRO resulting from the
break-up of the nanowires.  The observed differences from the two processing conditions can be
explained on the basis of the various mechanisms of fluid motion operating under the two irradiation
conditions. When irradiating the metal film with a 2-beam interference pattern, the resulting
periodic laser intensity induces a transient and periodic thermal gradient along the plane of
the film which creates a surface tension gradient not present in the single beam irradiation.
Consequently, $TC$ or Marangoni convection of the molten Co can occur. To contrast the results
of single beam and 2-beam irradiation, the time scales for the various mechanisms were estimated.
The timescale for $TC$ flow can be expressed as \cite{Schwarz-Selinger}:\begin{equation}
\tau_{Ma}=\frac{\Lambda_{laser}^{2}\eta}{4\frac{\partial\gamma}{\partial T}\Delta Th_{0}}\label{eq:TCtimescale}\end{equation}
where $\Lambda_{laser}=350\, nm$ is the laser fringe spacing, $\eta=4.45\times10^{-3}\, Pas$,
is the Co metal viscosity, $\frac{\partial\gamma}{\partial T}=-0.34\times10^{-3}Jm^{-2}K^{-1}$
is the rate at which the surface tension of the Co changes with temperature and $\Delta T$ is
the maximum temperature difference between the peak and valley of the laser fringe.  The timescale
for the TFH instability can be expressed as \cite{vrij,vrij2}: \begin{equation}
\tau_{s}=\frac{96\pi\gamma\eta}{A^{2}}h_{0}^{5}\label{eq:SpinTimescale}\end{equation}
An important aspect of the above equations is that the time scale of the $TC$ flow \emph{decreases}
with increasing $h_{0}$ while the TFH instability \emph{increases} with increasing $h_{0}$.
 We determined that typical liquid lifetimes range from $1\leq\tau_{L}\leq10\, ns$ and thermal
gradients range from $0.5-0.7\,\frac{K}{nm}$ \cite{favazza06c,Trice06a}. From Eq. \ref{eq:TCtimescale}
and \ref{eq:SpinTimescale}, $\tau_{s}>\tau_{Ma}$ for films with $h_{o}\leq2\, nm$ and $\tau_{s}<\tau_{Ma}$
for films with $h_{0}>2\, nm$. This implies that pattern formation is dominated by the $TC$
flow for $h_{0}>2\, nm$ with the TFH flow dominating for $h_{0}<2\, nm$, as shown in the comparative
timescale plot in Fig. \ref{fig:TimescalesandImages}. This result was confirmed experimentally
as nanowire formation was observed only above $2\, nm$ while it was absent for films $<2\, nm$,
as shown in Fig. \ref{fig:TimescalesandImages}(inset A). For film thicknesses near the cross-over
point of $h_{0}=2\, nm$, the pattern consisted of particle formation with some evidence for
lateral movement of the metal (Fig. \ref{fig:TimescalesandImages}(inset B)) indicating that
both mechanisms are operative \cite{moriarty02}. 

For films with $h_{0}>2\, nm$ nanowire formation under the 2-beam irradiation also permitted
access to a Rayleigh-like instability in which cylinders are unstable to wavelengths $\ge2\pi r_{cyl}$,
with the fastest growing wavelength scaling as $\Lambda_{R}\propto r_{cyl}$ , where $r_{cyl}$
is the radius of the cylinder. The characteristic time scale for this process can be expressed
as \cite{deGennes,rayleigh}:\begin{equation}
\tau_{R}=\sqrt{\frac{\rho r_{cyl}^{3}}{\gamma}}\label{eq:RayleighTimescale}\end{equation}
where $\rho=7.8\,\frac{g}{cm^{3}}$ is the liquid density of Co \cite{Smithells}. Assuming these
are partial cylinders with contact angles $>{90}^{\circ}$and using the projected width of these
regions from SEM images, we estimated  $r_{cyl}$, preceding the break-up into nanoparticles,
to be in the range of $35-55\, nm$.  Based on these values of $r_{cyl},$ the typical magnitude
for the cylinder break-up time was estimated as $\tau_{Ray}\sim1-2\, ns$. Since this time scale
is comparable to or smaller than the typical liquid lifetimes, the Rayleigh process is clearly
accessible whenever a cylinder is formed. Also, according to the Rayleigh formulation, the particle
radius scales linearly with both the spacing between particles and the radius of the original
cylinder \cite{rayleigh,rayleigh2,eggers}. Through volume conservation,  the cylinder radius
varies as $r_{cyl}\propto h_{o}^{1/2}$, resulting in a  similar  scaling relation for the size
of the particle as $r_{p2}\propto h_{o}^{1/2}$. As shown in Fig. \ref{fig:LengthscalePlots}(a-b),
the particle spacing and the size both scale as $h_{o}^{1/2}$ and the ratio of particle spacing
to radius is independent of film thickness. While,  the ratio in the classical Rayleigh break-up
of a perfect cylinder is expected to be $\frac{\lambda_{NN2}}{r_{p2}}\sim4.7$ \cite{rayleigh,rayleigh2,eggers}
our measured ratio was $\simeq5.6$. We have ruled out previously existing explanations for how
this ratio can be modified \cite{nichols,mccallum,hocking,hocking2,davis,gurski}. Nichols et
al. \cite{nichols} have shown that if the break-up is dominated by solid-state surface diffusion,
then the ratio is maintained,  however, if the flow is generated by external volume diffusion,
then the ratio grows to $\sim6.1$. However, the timescale over which external volume diffusion
can cause break-up has been estimated as $\sim0.3\, s$ ($\tau_{vol}\sim\frac{13D_{V}\gamma\Omega}{fkT}$,
where $D_{v}$ is the volume self-diffusion coefficient, $\Omega$ is the atomic volume and $f$
is the correlation factor \cite{nichols}) for cylinder with $r_{cyl}=40\, nm$. This timescale
is  well outside the realm of the laser processing time (for $t=0.3\, s,\; n\sim O(10^{6})$),
and as such can be disregarded as a possible break-up mechanism. We have also estimated the correction
to the ratio for situations in which truncated cylinders break-up based on the work by McCallum
et al. \cite{mccallum} and determined that only a small change to the ratio will be introduced.
 While there have also been papers examining the effect of having a non-uniform contact line
or an anisotropic surface, these models also do not yield the observed 5.6 ratio \cite{davis,gurski}.
It is possible that the thermocapillary flow and a Rayleigh-like instability can drive self-organization
of the film such that the liquid profiles are stationary solutions to the non-linear equations
describing the pattern evolution \cite{Mitlin93,Tan}. However, a nonlinear analysis along with
further experiments are required to verify such a hypothesis. 

In conclusion, self-organization leading to spatially correlated nanostructures under ns laser
irradiation of ultrathin Co films on SiO$_{\text{2}}$ has been investigated. For films with
thickness $2-8\, nm$, non-uniform laser irradiation by 2-beam interference leads to pattern
formation characterized by the formation of nanowires via thermocapillary flow and eventually
to nanoparticles via a Rayleigh-like break-up of the nanowires. The nanowire break-up results
in an average nearest-neighbor particle spacing of $\lambda_{NN2}\propto h_{0}^{1/2}$ with radius
$r_{p2}\propto h_{0}^{1/2}$ in contrast to the TFH instability in which $\lambda_{NN1}\propto h_{0}^{2}$
and $r_{p1}\propto h_{0}^{5/3}$. For films with $h_{0}\leq2\, nm$ nanowire formation was absent
because the TFH time scales were much shorter than the $TC$ timescale.  These results show that
self-organization through laser-induced hydrodynamic flow can be used to make a variety of strongly
correlated surface nanostructures on amorphous substrates by selecting the appropriate film thickness
based upon thermophysical parameters. 

RK and RS acknowledge support by the NSF through a CAREER grant  DMI-0449258 and grant CTS 0335348
respectively. \textbf{\large }{\large \par}

{*}Electronic address: ramkik@wuphys.wustl.edu, $^{\dagger}$Electronic address: suresh@wustl.edu


\pagebreak

\section*{Figure captions}

\begin{figure}[H]
\begin{centering}\subfigure[]{\includegraphics[height=2.5in,keepaspectratio]{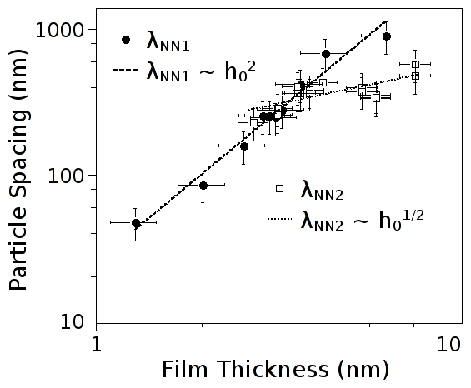}}~\subfigure[]{\includegraphics[height=2.5in,keepaspectratio]{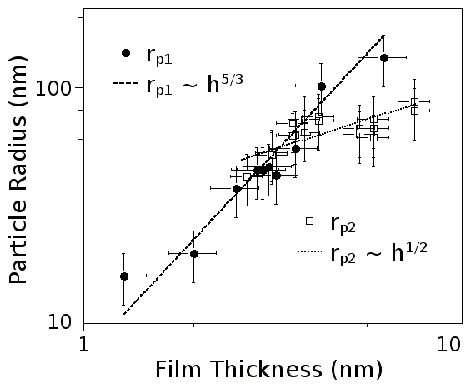}}\par\end{centering}

\caption{\label{fig:LengthscalePlots}}
\end{figure}

\begin{figure}[H]
\begin{centering}\subfigure[]{\includegraphics[height=1.75in,keepaspectratio]{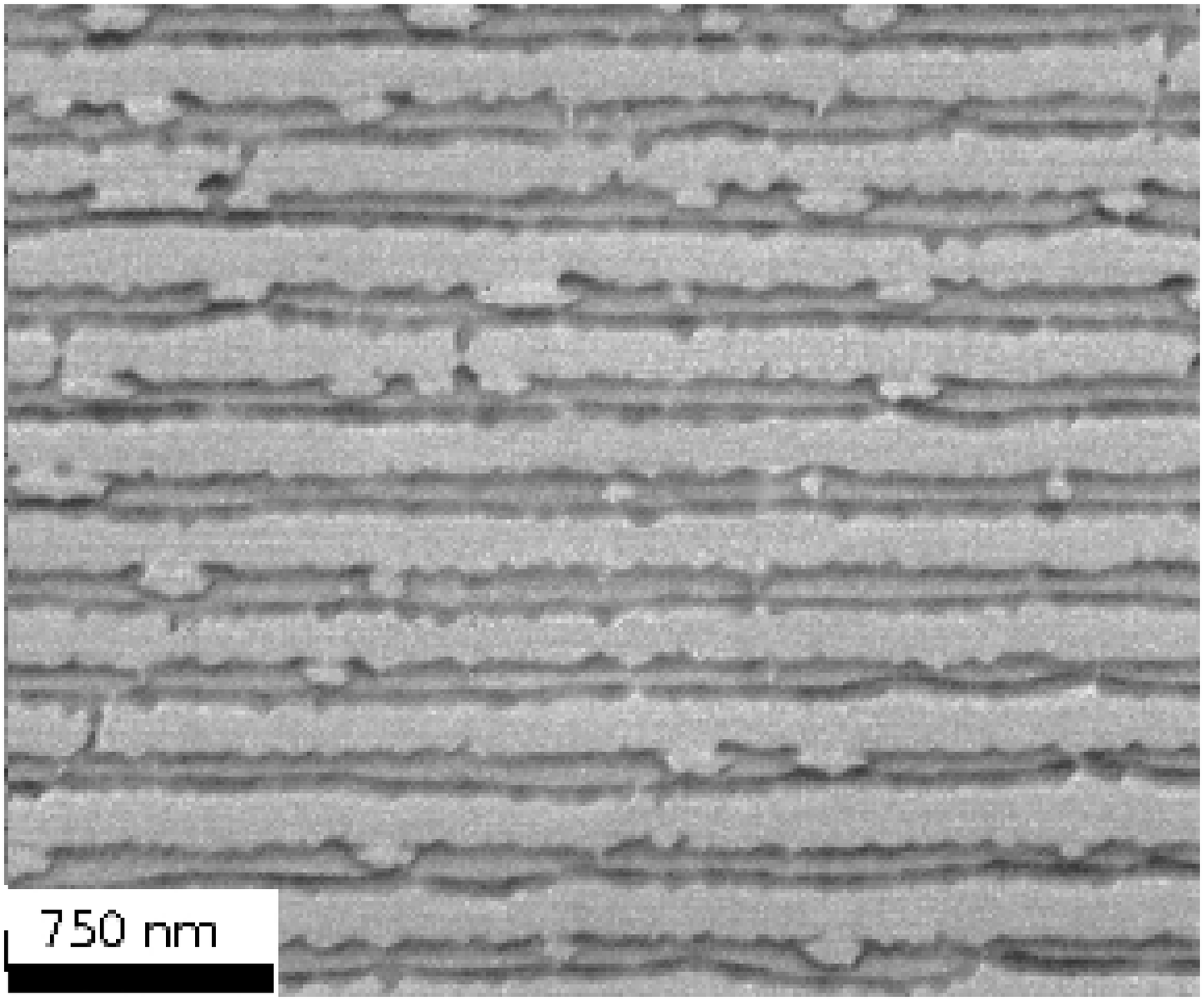}}\subfigure[]{\includegraphics[height=1.75in,keepaspectratio]{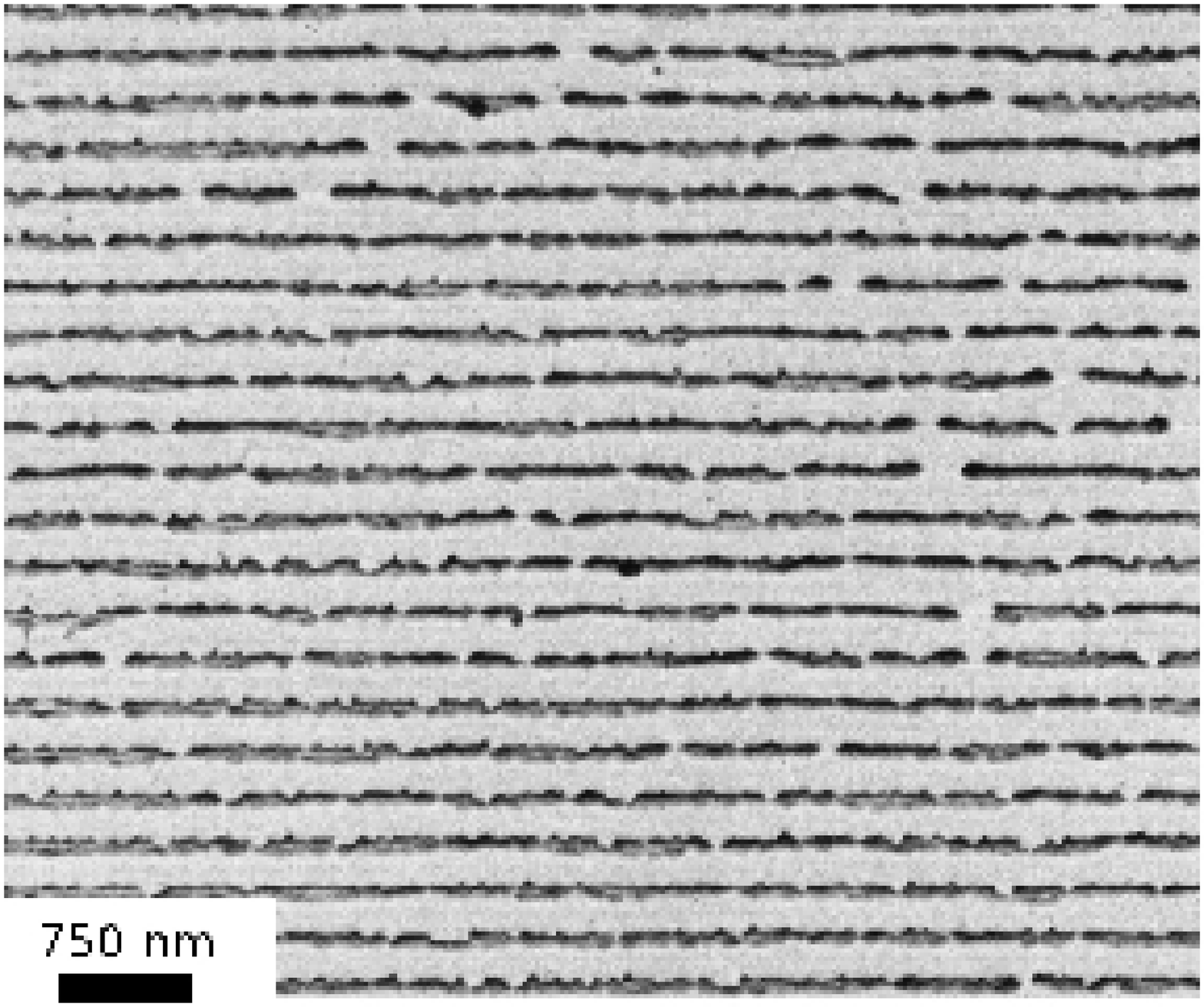}}\subfigure[]{\includegraphics[height=1.75in,keepaspectratio]{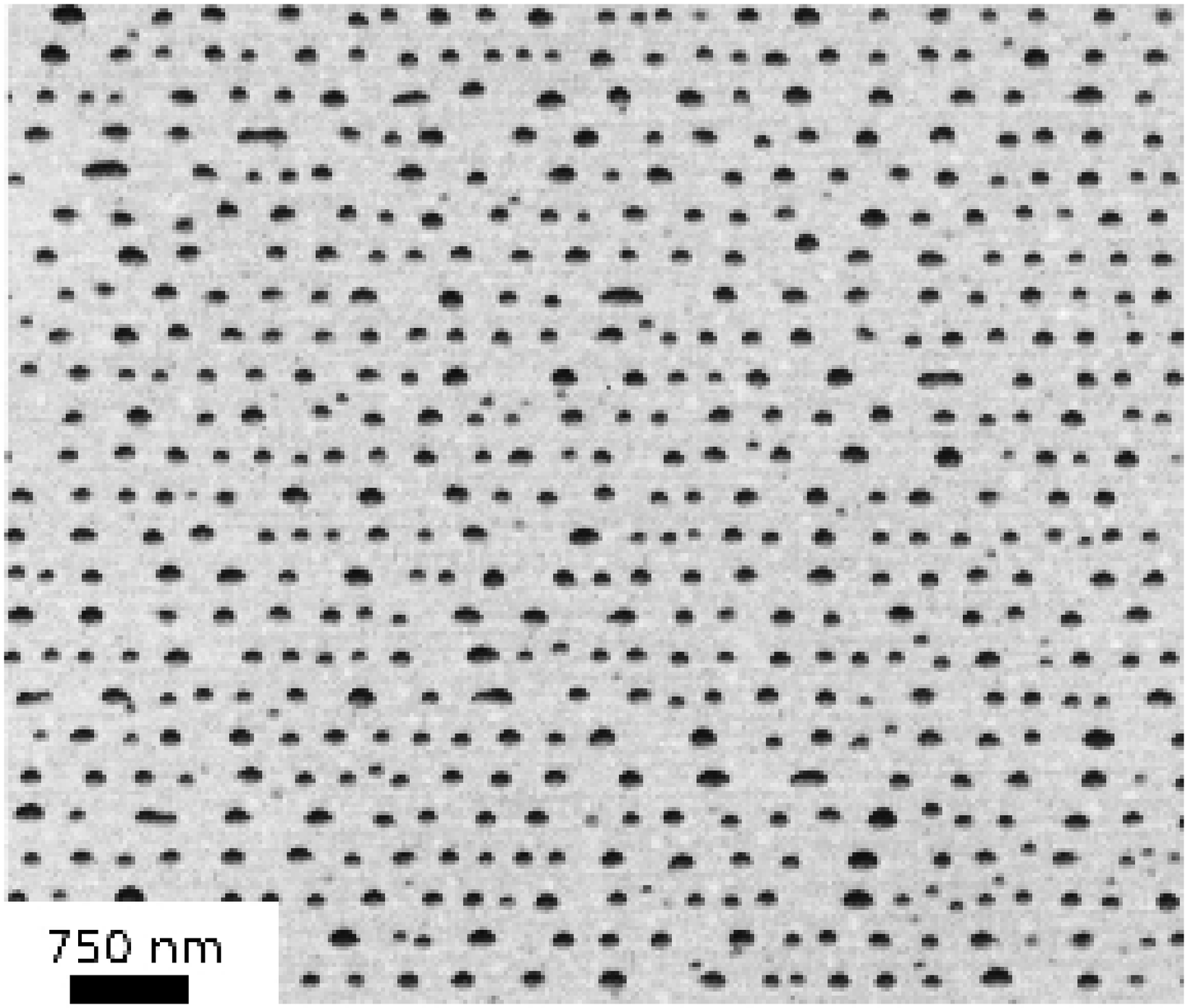}}\par\end{centering}

\caption{\label{fig:PatternProgression}}
\end{figure}

\begin{figure}[H]
\begin{centering}\includegraphics[height=3in,keepaspectratio]{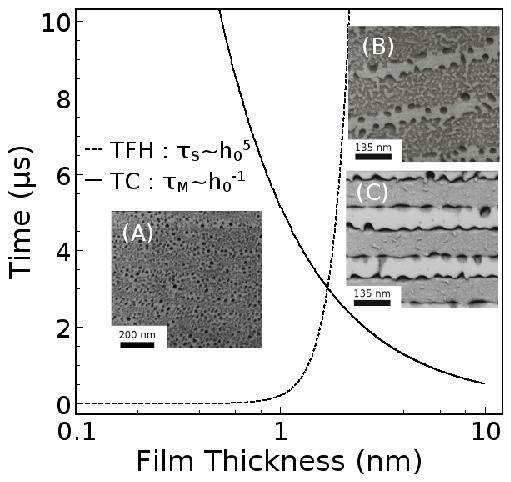}\par\end{centering}

\caption{\label{fig:TimescalesandImages} }
\end{figure}

\begin{itemize}
\item Figure \ref{fig:LengthscalePlots}: Self-organized length scales versus  initial film thickness:
(a)  $\sqrt{h_{0}}$ dependency for the particle radius from 2-beam interference irradiation
(open squares); and $h_{0}^{\frac{5}{3}}$ dependency from single beam irradiation (closed triangles).
(b)  $\sqrt{h_{0}}$ dependency for the particle spacing from 2-beam irradiation (open squares)
and $h_{0}^{2}$ dependency from single beam irradiation (closed triangles).
\item Figure \ref{fig:PatternProgression}: SEM images depicting the stages of pattern formation for
2-beam interference irradiation of a $\sim6$ nm film as a function of increasing number of laser
pulses $n$: (a) periodic rupture (b) nanowires and (c) the final nanoparticle state which exhibits
both LRO and SRO. (Co rich and SiO$_{\text{2}}$ rich regions correspond to bright and dark contrast
respectively). 
\item Figure \ref{fig:TimescalesandImages}: Comparison of the time scales for TFH dewetting and thermocapillary
($TC$) flow as a function of initial film thickness with the morphology for various film thickness
following irradiation (insets). (A) $h\le1\, nm$ film showing that the entire film dewets; (B)
$h\sim2\, nm$ film showing that while $TC$ flow causes the film to split there is still evidence
for TFH dewetting and (C) $h\sim4.5\, nm$ film showing that $TC$ forces completely dominate
morphology change. (Co rich and SiO$_{\text{2}}$ rich regions correspond to bright and dark
contrast respectively). 
\end{itemize}
\pagebreak

\pagebreak

\pagebreak

\end{document}